# Feature Evolution and Reuse - An Exploratory Study of Eclipse

Amjed Tahir
School of Eng. and Advanced Technology
Massey University, New Zealand
a.tahir@massey.ac.nz

Sherlock A. Licorish and Stephen G. MacDonell
Department of Information Science
University of Otago, New Zealand
{sherlock.licorish;stephen.macdonell}@otago.ac.nz

**Abstract**

*One of the purported ways to increase productivity and reduce development time is to reuse existing features and modules. If reuse is adopted, logically then, it will have a direct impact on a system's evolution. However, the evidence in the literature is not clear on the extent to which reuse is practiced in real-world projects, nor how it is practiced. In this paper we report the results of an investigation of reuse and evolution of software features in one of the largest open-source ecosystems - Eclipse. Eclipse provides a leading example of how a system can grow dramatically in size and number of features while maintaining its quality. Our results demonstrate the extent of feature reuse and evolution and also patterns of reuse across ten different Eclipse releases (from Europa to Neon).*

## 1. INTRODUCTION

Reuse has long been promoted in the literature on software development as a way to improve the quality of both software process and product [1]. Several software development methods have embraced the concept of reuse as a key element of their processes: modern software engineering approaches such as Component-Based Software Engineering (CBSE) and Software Product Lines (SPL) consider reuse (and the reusability of components) as fundamental to their activities. One of the main goals of such approaches is to build new software components that can be easily reused and integrated with other (new or existing) software components.

Many contemporary software systems depend to varying extents on reuse thanks to the open source movement, which has made a great deal of reusable software available in the public domain. In fact, reuse of source code or libraries has been identified as one of the success factors of open source projects [2]. Projects resulting in Linux OS, Apache Software, and Eclipse demonstrate how open source libraries have been used and expanded by other projects. Reuse of open source software components within commercial software can also be crucial, as many commercial software organizations would simply not be able to function without some open source software [3]. Evidence on just *how* components are reused, and the relationship between software reuse and system evolution, is limited, however. It is thus of interest to understand how such systems evolve and the specific features that are reused. Here we define a feature as a prominent or distinctive user- visible aspect of a software product or service [4].

We provide initial evidence around how software systems evolve in this preliminary work by exploring versions of the popular Eclipse system, with a focus on feature evolution and reuse. Our contributions are twofold. We provide insights into how features and packages are reused during project evolution, adding to the body of evidence around software evolution. We also provide practical suggestions for how our findings may be useful for software development practitioners.

The remaining sections of the paper are organized as follows. In Section 2 we review related work, and we describe our study setting in Section 3. Our results are presented in Section 4, and we discuss these results and evaluate their implications in Section 5. We then consider threats to the work's validity in Section 6, before providing concluding remarks in Section 7.

## 2. RELATED WORK

The study presented here sits at the intersection of software reuse and software evolution. Both areas of research have attracted substantial interest and attention, aimed at providing insights into the way software changes over time.

Early work by Basili et al. [5] evaluated the impact of software reuse on the development of OO systems, finding that reuse has a strong positive impact on productivity and quality. These authors further revealed that reusable software components are associated with reduced defect density. Mockus [6] studied code reuse in multiple open source projects and found that half of the 5.3 million files studied were used in more than one project. The most widely reused set of files were language translations for user messages followed by install modules for Perl. Some of the widely reused components involved a set of hundreds of files. Haefliger et al. [7] studied the extent of source code reuse in open source projects, using quantitative data (through mining source code repositories) and qualitative data (through interviews with developers) gathered from a sample of six medium to large open source projects. The main finding of their study showed that source code reuse was quite extensive across the sample of projects studied, with developers actively reusing available code and other knowledge that solved their technical problems.

Software evolution as a topic of research has also received noteworthy attention. Asaduzzaman et al. [8] studied the specific changes that introduce bugs in Android over time and found that larger code changes were likely to result in more defects. artifacts were used by Thomas et al. [9] who applied topic modelling to study the evolution of JHotDraw and JEdit, finding correlations between the evolution of various topics and developers code changes. Zaidman et al. [10] also used artifacts and the TeMo tool to evaluate the co-evolution of test and production code. From an end-user perspective, Licorish [11] examined the evolution of architecture issues and non- functional requirements in the Android project finding that most architecture-related issues were located in the Android application layer and usability-related concerns were reported most when they were held to be given greatest attention. Further, in an examination of software repository data, Herraiz [12] observed that a small subset of metrics is sufficient to characterize a software system, and software evolution is generally a short-range correlated process. Further works have considered various other aspects of software evolution, including evolutionary code complexity [13], defect classification [14], origin analysis and refactoring [15], reuse [16], project communication [17] and teams' behavioral processes [18].

Recent work of Bakar et al. [19] proposed a new approach, known as Feature Extraction for Reuse of Natural Language requirements (FENL), to extract phrases that can represent software features from online user reviews (such as app reviews) to better understand requirements reuse. The focus of their work was on the development of a tool and the accuracy of the output compared to the other available approaches. Martinez et al. [20] [21] also proposed a feature identification and localization tool (BUT4Reuse) that extracts features from existing software packages. We explain our use of this tool in Section 3.

While previous works have examined various aspects of software reuse and systems evolution, we know very little about the survival rate of software features and packages resident is large systems. The software engineering community would benefit from insights into the specific properties of features that propagate over project evolution, as therein would be pointers for contributors of open source projects in terms of how their contributions would be most valuable. Insights into how feature reuse and system evolution materialize would also be useful for contributing evidence towards validating the laws of software evolution [22]. We thus develop a preliminary research question to study this issue across ten Eclipse releases:

***RQ: how do features evolve over different software releases?***

## 3. STUDY SETTING

The goal of this work is to investigate feature evolution and reuse, using the case of Eclipse[1] - one of the largest and most successful open source ecosystems. We consider Eclipse to be a very suitable artifact for this study as it enables us to study feature evolution and reuse at a relatively large scale. Eclipse is built on multiple frameworks that together form the Eclipse platform. Eclipse has undergone multiple releases (to date, the total number of official releases is 14), the first edition (Eclipse 3.0) having been released in 2004. The platform contains multiple tools, APIs and frameworks. While the Eclipse IDE supports development in many programming languages, including JavaScript, Haskell, Python, Scala and Rust, Eclipse is most well known for its support for Java development, and it is claimed to be the most widely used IDE by Java developers[2].

To study feature evolution and reuse in Eclipse we extracted features from ten Eclipse releases, from *Europa* (2007) through to *Neon* (2016). Eclipse publish their major versions (i.e., their main releases, which include Eclipse platform and associated projects) annually in June. Table I provides detailed information about the Eclipse releases that we included in this study. Note that we only included *major* and *stable* releases of Eclipse. We chose *Europa* as our starting release as this was the first version for which we managed to obtain its full packages. Before *Europa*, there were only three known stable releases (*3.0, 3.1 and 3.2 (Callisto)*) but we could not obtain all packages of these releases at the time we extracted the data in April 2016.

To identify and locate features in the Eclipse releases we used BUT4Reuse[3] [21], a feature location tool that extracts feature names and their plugin dependencies, using a range of techniques, from each Eclipse release. We carefully followed the instructions[4] provided by Martinez et al. [20]. An example of a relationship between features and plugins is shown in Table II.

First we obtained all package variants for all ten Eclipse releases that we studied from the Eclipse official site. For each release, we used the Windows 64-bit packages. The Eclipse website provides details of the main packages that are included in each release. We report the total number of packages comprising each release in Table I. BUT4Reuse takes the full project directory (holding the project's package variants including all plugins) as an input and produces a list of features and plugins and their relationships as an output. These can then be extracted as a separate CSV file. While the tool employs a number of feature location techniques that can be used to extract features and their dependencies describing them is beyond the scope of this paper: details of these techniques and the implementation of the tool are provided in Martinez et al. [20]. In brief, we used a combination of Strict Feature-Specific (SFS) and Term Frequency - inverse document frequency (TF-idf) techniques, as these appeared to be the most effective options of those available in terms of maximizing precision and recall.

---

[1] https://www.eclipse.org/
[2] http://tinyurl.com/y8alul9c
[3] https://but4reuse.github.io
[4] https://github.com/but4reuse/but4reuse/wiki/Benchmarks

TABLE I. ECLIPSE RELEASE INFORMATION

| Eclipse release | Release date | Version | #Packages | #Features | #Plugins mapped to features |
|---|---|---|---|---|---|
| Europa | 29/06/2007 | 3.3 | 4 | 91 | 499 |
| Ganymede | 25/06/2008 | 3.4 | 7 | 291 | 1413 |
| Galileo | 24/06/2009 | 3.5 | 10 | 341 | 1818 |
| Helios | 23/06/2010 | 3.6 | 11 | 305 | 1666 |
| Indigo | 22/06/2011 | 3.7 | 11 | 337 | 1896 |
| Juno | 27/06/2012 | 4.2 | 13 | 406 | 2224 |
| Kepler | 26/06/2013 | 4.3 | 12 | 437 | 2286 |
| Luna | 25/06/2014 | 4.4 | 13 | 548 | 2606 |
| Mars | 24/06/2015 | 4.5 | 12 | 549 | 2745 |
| Neon | 22/06/2016 | 4.6 | 14 | 567 | 2786 |

TABLE II. EXAMPLES OF FEATURES AND THEIR ASSOCIATED PLUGINS AND FRAMEWORKS FROM ECLIPSE

|  | Example 1 | Example 2 |
|---|---|---|
| Release | Luna | Ganymede |
| Feature name | Eclipse Product Configuration | EMF SDO Editor |
| plugin | org.eclipse.rcp.configuration | org.eclipse.emf.ecore.sdo.editor |
| Framework | Rich Client Platform | Eclipse Modeling Framework |

While precision was not very high (29% in the case of *Helios*), recall was relatively better (60% in the case of *Juno*). Our full data set (and the results extracted from all ten releases) are publicly available for future replication studies[5].

## 4. RESULTS

We explore various aspects of Eclipse features and how they have evolved over the study period. As explained in Section III, we consider feature evolution across ten consecutive Eclipse releases that span a decade, from *Europa* (version 3.3 - 2007) through to *Neon* (version 4.6 - 2016). We divide our results into two parts: the first describes a range of general evolution and reuse patterns, while the second explains more specific patterns of interest.

### A. General patterns

Our first observation is that the platform has grown dramatically in size over the study period. Apart from a few minor deviations there has been a generally steady increase in the number of features included (see Figure 1). The total number of features has in fact increased by a factor of 6, growing from 91 features in *Europa* to 567 in *Neon*. This is directly related to the increase in the number of packages (and, therefore, plugins) over the different releases (Figures 2 and 3) - although the *Helios* and *Juno* releases exhibit a slight decrease in the number of packages compared to the previous release. On average, the number of packages increased (between succeeding releases) by a factor of 1.17 whereas the number of features increased by a factor of 1.33.

We also investigated the presence and nature of singular features (i.e., features that appear only in one single release). Figure 4 shows the number of unique features in each of the ten releases. As is evident, the proportion of unique features increased in the *Ganymede* and *Galileo* releases before decreasing again in following releases.

We note that *Mars* (the second to last release considered) had the highest proportion of unique features (27% of the total number of features): the number of unique features had increased by a factor of 2.63, from 71 features in *Luna* to 187 features in *Mars*. We observed that the vast majority of these features are related to one specific framework: the Eclipse Modeling Framework (EMF)[6] (47% of the features). After filtering out all overlapping features we identified a total of 1637 unique features across the ten releases examined. By analyzing their usage patterns, we observed that the majority of features may be considered as reusable - 944 features (58% of the total number of features) were used in two or more releases. That said, some 693 features (42%) have not been reused (i.e., they appear in only one release). *Mars* and *Neon* contain the most 'single-release' features.

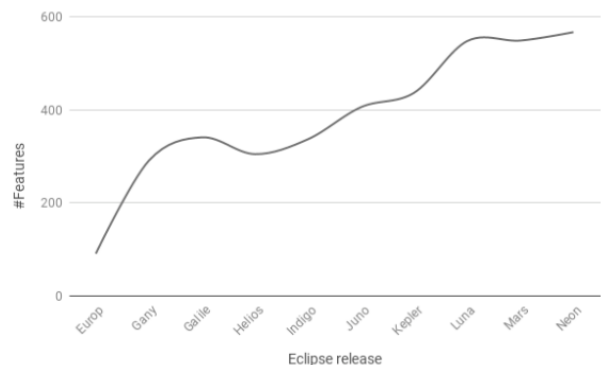

Fig. 1. Number of features across all releases

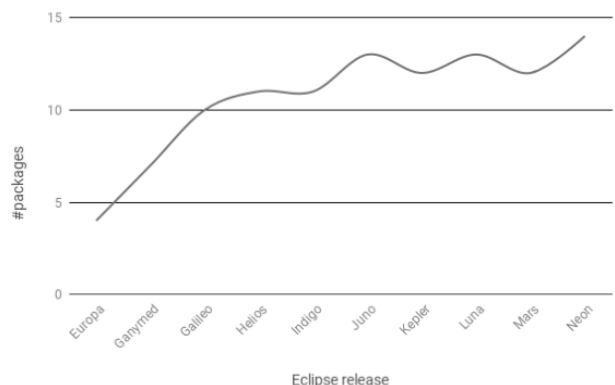

Fig. 2. Number of packages across all releases

We did not find any feature that appears in all ten releases studied. The highest number of releases that a feature appears in was eight, with *Kepler* and *Mars* being the releases that most features did not appear in. Only 3% of the features (49 out of 1637) were reused in 8 releases, with almost half of these features related to the Eclipse Data Tools Platform (DTP) Project[7]. Examples of such features are: "DTP Enablement for MySQL", "DTP Enablement for

---

[5] http://tinyurl.com/APSEC-17
[6] http://www.eclipse.org/modeling/emf

[7] https://eclipse.org/datatools

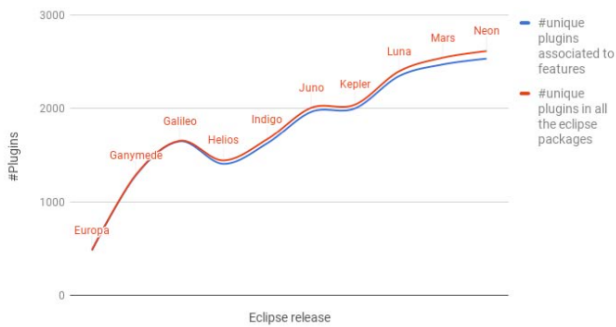

Fig. 3. Number of plugins across all releases

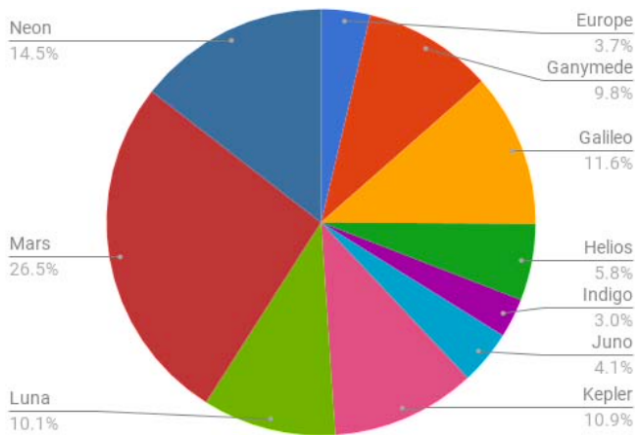

Fig. 4. Number of unique features in each release

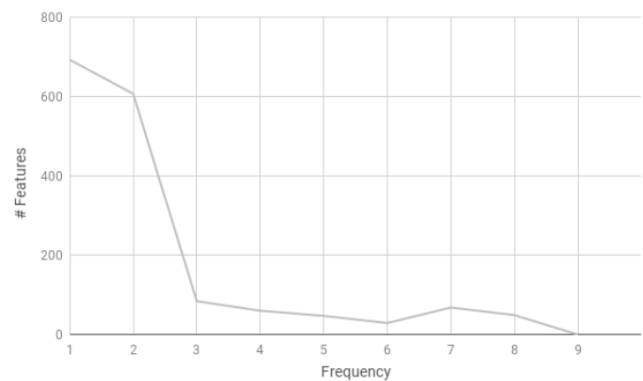

Fig. 5. How frequently does a feature appear in releases?

Oracle" and "Eclipse XML Editors and Tools". In terms of features that appear only in two releases, we observed that *Kepler* and *Mars* share the highest number of such features. There are 355 features that appear only in those two releases, even though these particular releases are not sequential (i.e., the *Luna* release occurred between them). A large portion of this commonality is related to the extension of functionality of specific frameworks and tools such as the EMF, Mylyn[8] and the Xtend Framework[9]. A similar pattern is evident between *Luna* and *Neon*. Examples of features that appear here are: "Xtext Complete SDK" and "Mylyn Tasks Connector: Bugzilla".

Figure 5 shows the frequency with which a feature appears (i.e, how frequently does a feature appear in releases?). While there are 944 'reused' features, there is also a large number of features that were not reused in those selected major releases. We had expected a higher number of reused features across all Eclipse releases. However, most of those reusable feature appear in 2 or 3 releases (691 features - around 74% of the total number of reused features).

**B. Specific patterns**

Most features that we observed in the data follow a specific pattern - reused features appear in sequential releases (from one release to the next one; for example, a feature appears in *Ganymede* and *Galileo*, or in *Indigo* and *Juno*). There are some specific exceptions to this pattern and they are mostly related to *Kepler* and *Mars*. In these two non-sequential releases most features either appear only in these two releases, or appear in all other releases except those two. While this is an interesting pattern we are unsure of the reasons for it. Furthermore, we found other features that do not conform to this specific pattern. For example, a feature that is related to the EMF framework known as "EMF Core Runtime" exists in *Ganymede*, is missing from the following release (*Galileo*) but appears again in the three subsequent releases.

Other features exhibit more intriguing patterns, as they exist in one release and then miss the next two or three releases, before appearing again in later releases. For example, a feature related to the C/C++ development kit known as "CDT GNU Toolchain Debug Support" existed in *Ganymede* and *Galileo*, then did not appear in the following four releases (between *Helios* and *Kepler*) before appearing again in *Luna* and *Neon*. Similarly, another feature related to the Dynamic Languages toolkit framework was included in the *Galileo* and *Helios* releases, was missing from the following three releases (*Indigo-Kepler*), before appearing again in *Luna* and *Neon*.

## 5. DISCUSSION AND IMPLICATIONS

*How do features evolve over different software releases?*

Taken as a whole, the Eclipse platform has grown dramatically in size over the years since its introduction. This has implications for new contributors joining the community in terms of their ability to come up to speed with the code base, but equally for the management of the project. We noticed that some features were only evident in specific releases, while there were times when features formed the core to the majority of releases. This pattern holds generally for software systems, which tend to have a stable core with various features developed around it to add specific value for users. That said, we did notice changes to the core in some releases. The fact that unique features come and go frequently could impact on contributors that may not be able to remain active continuously. Should such members withdraw from the community temporarily, their level of familiarity would diminish with a rapidly changing system. This could be detrimental for those dependent on the open source community [3], where the expectation of feature-set stability may be violated.

In fact, most features that we observed in the Eclipse releases follow a specific pattern - reused features tend to appear in sequential releases (from one release to the next one). This could be a deliberate strategy related to the cycle of change for Eclipse versions. Perhaps such a strategy also provides the user community with an opportunity to evaluate features for a sufficiently long period of time. Exceptions to this norm were also noticed for a few features, and particularly in terms of features that missed successive releases before returning. There is little doubt that such features may return due to community demands, but may equally resurface because of their relevance in the code base's overall stability. Findings here support the relevance of reuse and software evolution studies [9]. Such studies are key to understanding how soft- ware systems evolve.

---

[8] http://www.eclipse.org/mylyn

[9] http://www.eclipse.org/xtend

Our findings here also somewhat conform to those emphasized by Lehman [22]. He noted that systems exhibited some level of instability or chaotic behavior after some specific releases, with much smoother trends between others. This pattern is observed in the figures above, and Figure 3 in particular. Lehman also noted that systems will go into decline unless they are rigorously maintained and adapted to operational environment changes. Key to the environment change mentioned here is user feedback [11], which offers an avenue for us to conduct follow-up work to explain the patterns uncovered in this work. This work encourages software engineering researchers and practitioners to carefully investigate the development of new plugins, their evolution and possible reuse patterns, as highlighted in this study.

## 6. THREATS TO VALIDITY

*Tool precision:* the main threat to the validity of this study relates to the use of the selected tool, BUT4Reuse, to extract features from Eclipse. The tool provides a wide range of feature location algorithms that use techniques such as natural language processing (NLP) and Information Retrieval (IR) to extract and locate features in a project. However, the reported overall reliability of the tool (in terms of precision and recall) shows that it might not be able to correctly identify some of the features. While this is a problem (especially in case we miss some important features), our aim was to provide a general observation of feature patterns and usage in Eclipse, and features were consistently extracted across versions (i.e., if the tool identified a feature in the first version it would be capable of doing so for that specific feature across all versions). A technique that can provide better precision will certainly help to complement our analysis. It would also be interesting to replicate the Eclipse case with other feature location techniques to see if the same patterns are observed.

*Eclipse data:* the data we obtained from Eclipse contain features from ten different major (and stable) releases. However, our data does not take into account sub-releases or minor releases (e.g., different builds). By including minor releases we might be able to explain some of the feature evolution patterns that we could not detect in major releases. For example, a certain feature might appear in a particular minor release, but then disappear in major releases. We believe that obtaining data from all major and minor releases will provide a fuller understanding of feature evolution and reuse.

## 7. CONCLUSIONS AND FUTURE WORK

This study investigated feature evolution and reuse in Eclipse, one of the most widely used IDEs. We extracted features from ten successive Eclipse releases (from *Europa* to *Neon*) using BUT4Reuse, a feature identification and localization tool, using two feature localization techniques (SFS and TF-idf). Based on our analysis of the collected data we observed that feature use and reuse in Eclipse projects has evolved over the study period. This appears to be the result of the evolution of the system's size and also the number of frameworks and plugins that are associated with the system. In terms of the evolution of features and packages, we found that there has been a generally steady level of increase in the number of packages, features and plugins between 2007 and 2016.

Out of a total of 1637 unique features, we found that the slight majority (58%) were reusable (that is, they have been used in two or more different releases). However, the other 42% of features have not been reused (i.e., they are features that appeared in only one release), with the *Mars* and *Neon* releases containing the most 'single-release' features. We also found that there are multiple reuse patterns that appear in all Eclipse releases.

Our findings provide support for previous evidence around system evolution. We also believe that the findings here may have implications for the availability of contributors with relevant skills and awareness in light of potential changes in the software, should they become temporarily inactive in the community.

That said, our work is preliminary, and thus there are issues that need further investigation. For future studies, a finer-grained analysis of minor releases would provide additional understanding for what actually happens between major releases. Incorporating information from minor releases might explain some of the evolution and reuse patterns that we could not detect by considering only major releases. In addition, using project development history information (such as information mined from version control and issue tracking systems) to see if details evident in these systems correlate with the reuse and evolution patterns we observed (and any recorded decisions related to reuse) would be useful. Another future research opportunity would be to compare the evolution and reuse patterns that we found in Eclipse with those found in other well-established open-source ecosystems such as the Apache[10] and Mozilla[11] projects.

---

[10] https://www.apache.org

[11] https://www.mozilla.org